\documentclass[aps,preprint,prd,showpacs,nofootinbib]{revtex4}
\usepackage{latexsym}
\usepackage{amsmath}
\usepackage{graphicx}
\usepackage{subfigure}
\usepackage{dcolumn}
\usepackage{bm}
\usepackage{amssymb}
\usepackage{latexsym}
\usepackage{ulem}
\usepackage{color}
\usepackage[colorlinks,linkcolor=magenta,anchorcolor=cyan,citecolor=blue]{hyperref}

\def\be{\begin{equation}}
\def\ee{\end{equation}}
\def\ba{\begin{eqnarray}}
\def\ea{\end{eqnarray}}

\def\nn{\nonumber}
\def\lf{\left}
\def\rt{\right}

\begin{document}

\title{Probing the primordial universe with gravitational waves detectors }

\author{Yu-Tong Wang$^{1}$\footnote{wangyutong12@mails.ucas.ac.cn}}
\author{Yong Cai$^{1}$\footnote{caiyong13@mails.ucas.ac.cn}}
\author{Zhi-Guo Liu$^{1}$\footnote{liuzhiguo@ucas.ac.cn}}
\author{Yun-Song Piao$^{1,2}$\footnote{yspiao@ucas.ac.cn}}

\affiliation{$^1$ School of Physics, University of Chinese Academy of
Sciences, Beijing 100049, China}

\affiliation{$^2$ Institute of Theoretical Physics, Chinese
Academy of Sciences, P.O. Box 2735, Beijing 100190, China}

\begin{abstract}

{ The spectrum of primordial gravitational waves (GWs),
especially its tilt $n_T$, carries significant information about
the primordial universe. Combining recent aLIGO and
Planck2015+BK14 data, we find that the current limit is
$n_T=0.016^{+0.614}_{-0.989}$ at 95\% C.L. We also estimate the
impacts of Einstein Telescope and LISA on constraining $n_T$.
Moreover, based on the effective field theory of cosmological
perturbations, we make an attempt to confront some models of early
universe scenarios, which produce blue-tilted GWs spectrum
($n_T>0$), with the corresponding datasets.  }

\end{abstract}

\maketitle

\section{Introduction}

In the recent years, searching for the primordial gravitational
waves (GWs) \cite{Starobinsky:1979ty, Rubakov:1982} has
invigorated the cosmological community. Since the primordial GWs
carry rich information about the early universe and the
UV-complete gravity theory, its detection would deepen our
understanding for the early universe scenario, see
Refs.\cite{Maggiore:1999vm,Guzzetti:2016mkm,Bartolo:2016ami}
for reviews.

The primordial GWs background, described mainly by its tilt $n_T$
and the amplitude $A_T=rA_s$ at the pivot scale $k_*$, spans a
broad frequency-band, $10^{-18}-10^{10}$ Hz. The slow-roll
inflation \cite{Linde:1981mu, Albrecht:1982wi} predicts a
slightly red-tilted spectrum with $n_T\simeq
-2\epsilon_{inf} <0$ and $0<\epsilon_{inf}\ll 1$. However, $n_T>0$
is interesting, since it boosts the primordial GWs background at
the frequency-band of laser interferometers (Advanced
LIGO/Virgo, as well as the space-based detectors),
e.g.\cite{Cai:2016ldn, Cai:2015yza}.

The primordial GWs at ultra-low frequency $10^{-18}-10^{-16}$ Hz,
or on large scale, may induce the B-mode polarization in the
cosmological microwave background (CMB)\cite{Kamionkowski:1996zd,Kamionkowski:1996ks}. The joint analysis of
BICEP2/Keck Array and Planck (BKP) data have put the constraint on
the amplitude of primordial GWs, $r_{0.05}<0.12$ (95\% C.L.)
\cite{Ade:2015tva}. Recently, the combination of above data and
Keck Array's 95 GHz data have improved the constraint to
$r_{0.05}<0.07$ (95\% C.L.) \cite{Array:2015xqh} (BK14). However,
no strong limit was found for the tilt $n_T$, which is because the
CMB band is too narrow to depict the whole property of GWs
background\cite{Zhao:2009mj,Huang:2015gca}.

Recently, the LIGO Scientific Collaboration, using ground-based
laser interferometer, has observed a GW signal (GW150914) with a
significance in excess of 5.1$\sigma$ \cite{Abbott:2016blz}, which
is consistent with an event of the binary black hole coalescence
based on general relativity (GR). Advanced LIGO (aLIGO) O1 put an
upper limit for the stochastic GWs background at the frequency
band $f\simeq 30\,{\rm Hz}$
\cite{TheLIGOScientific:2016wyq, TheLIGOScientific:2016dpb}.

The spectrum of primordial GWs is not only determined by the
evolution of the background in early universe scenarios,
e.g.\cite{Liu:2014tda, Li:2016awk}, but also
significantly affected by the modifications to GR. The Effective
Field Theory (EFT)
\cite{Cheung:2007st,Weinberg:2008hq,Gubitosi:2012hu,Gleyzes:2013ooa,Piazza:2013coa}
of cosmological perturbations offers a unifying platform to deal
with the cosmological perturbations of modified gravity theories,
such as the Horndeski theory \cite{Horndeski:1974wa} and its
beyond \cite{Gleyzes:2014dya}. The EFT method has
also been used in building healthy nonsingular cosmological
models \cite{Cai:2016thi, Creminelli:2016zwa}. Different
models of early universe scenarios generally have different predictions
 for $n_T$. Only when the measurements at different
frequency bands are combined, can one put tighter constraint on
the tilt $n_T$
\cite{Meerburg:2015zua,Cabass:2015jwe,Lasky:2015lej,Smith:2005mm,Smith:2008pf}.

Therefore, it is interesting to perform a joint analysis of recent
aLIGO O1 and Planck 2015+BK14 dataset to constrain $n_T$, as well
as confront the EFT parameters in corresponding models with the
data.

The planned space-based detector LISA will search the GWs signals
at the frequency about 1 mHz \cite{Seoane:2013qna, Ricciardone:2016ddg}
collect data in 2030s, and recently LISA Pathfinder has been
launched which paving the way for the LISA mission. While
the space-based GWs detection project also has been approved in
China. Moreover, the third generation ground-based detector,
Einstein Telescope (ET) \cite{Hild:2010id}, is also being planned,
which has higher sensitivity than aLIGO/VIRGO. Thus it will
also be interesting to estimate the capability of LISA
and ET in constraining the tilt $n_T$.

This paper is organized as follows. In Sec.\ref{Sec:SecII},
combining recent aLIGO and Planck2015 +BK14 data, we provide an
updated constraint on the tilt $n_T$ of primordial GWs, and
also forecast the impacts of LISA and ET. In Sec.\ref{Sec:SecIII},
based on EFT, we explore how to confront the models of early
universe scenarios, which produce blue-tilted GWs spectrum
($n_T>0$), with the corresponding datasets. Sec.\ref{Sec:SecIV}
is the conclusion.

\section{Method, results and forecasts}
\label{Sec:SecII}

\subsection{Upper limits put by interferometers}
\label{Sec:SecIIA}

Conventionally, one define
\begin{equation}
\label{density} \Omega_{\text{GW}}(k,
\tau_{0})=\frac{1}{\rho_{\text{c}}}\frac{d\rho_{\text{GW}}}{d\ln
k}=\frac{k^{2}}{12 a_0^2H^2_0}P_{T}(k)T^2(k,\tau_{0})\,
\end{equation}
to depict the relic GWs background, where
$\rho_{\text{c}}=3H^{2}_0/\big(8\pi G\big)$ is the critical
density, $\Omega_{GW}(k, \tau_{0})$ reflects the fraction of
$\rho_{\text{GW}}$ per logarithmic $k$-interval, and
$\rho_{\text{GW}}$ is the present energy density of GWs. The
primordial GWs spectrum $P_T(k)$ is generally
\begin{equation} P_T(k)=rA_s\left(\frac{k}{k_*}\right)^{n_T},
\,\label{eq:tensor_pk}
\end{equation}
where $n_T$ is the tilt of spectrum, and $A_s$ is the amplitude of
primordial scalar perturbations at the pivot scale $k_*$. When
applying Eq.(\ref{eq:tensor_pk}) to fit CMB data, usually one set
$k_*=k_{CMB}\sim 0.01\, \mathrm{Mpc}^{-1}$. Current constraint put by CMB
data is $r_{0.05}<0.07$ (95\% C.L.) with $k_*=0.05 \,\mathrm{Mpc}^{-1}$
\cite{Array:2015xqh} (BK14), which corresponds to
$\Omega_{\text{GW}}<10^{-15}$ at $k\sim 10^{-16}$ Hz. Transfer
function $T(k,\tau_0)$ is
\cite{Turner:1993vb,Zhao:2006mm,Zhao:2011bg,Zhao:2013bba}, \be
T(k,\tau_{0})=\frac{3
\Omega_{\text{m}}j_1(k\tau_0)}{k\tau_{0}}\sqrt{1.0+1.36\frac{k}{k_{\text{eq}}}+2.50\big(\frac{k}{k_{\text{eq}}})^{2}},
\label{Tk} \ee where $\Omega_{\text{m}}={\rho_{\text{m}}/
\rho_c}$, see also \cite{Kuroyanagi:2014nba}.

%We have neglected the effects of the different thermal histories.

Current upper bound put by aLIGO O1 is $\Omega_\mathrm{GW}<{
10^{-7}}$
\cite{TheLIGOScientific:2016wyq, TheLIGOScientific:2016dpb}
at the frequency $f\simeq 30\,{\rm Hz}$, in which the wavenumber
$k$ is related to $f$ by $f= {k}/{2\pi a_0}$. The network of aLIGO
and Virgo detectors operating at their final observing runs O5
will be able to arrive at $\Omega_\mathrm{GW}\sim 10^{-9}$.

%In detail, the low-frequency acceleration noise corresponding this
%configuration is given as
%\begin{equation}
%  S_{n,\rm acc}(f) = 9\times10^{-30} \frac{1}{(2\pi f)^4}\left(1+\frac{10^{-4}{\rm
%        Hz}}{f}\right)\,\rm{m}^2\,{\rm Hz}^{-1} ,\label{eq:acc}
%\end{equation}
%and the shot noise component can be represented by
%\begin{equation}
%  S_{n,\rm sn}(f) = 2.2\times10^{-23} \rm{m}^2\,{\rm Hz}^{-1} ,\label{eq:sn}
%\end{equation}

\subsection{Combined datasets (interferometers and CMB) and results}

The parameters set of the lensed-$\Lambda$CDM model is
$\{\Omega_bh^2, \Omega_ch^2, 100\theta_\mathrm{MC}, \tau,
\ln(10^{10}A_s), n_s\}$, with the baryon density $\Omega_bh^2$, the
cold dark matter density $\Omega_ch^2$, the angular size
$\theta_\mathrm{MC}$ of the sound horizon at decoupling, the
reionization optical depth $\tau$, the amplitude $\ln(10^{10}A_s)$
and the tilt $n_s$ of the primordial scalar perturbation spectrum.
We also include the parameters $r=A_T/A_s$ and $n_T$, which
satisfy (\ref{eq:tensor_pk}). We use the pivot scale
$k_*=0.01\,\mathrm{Mpc}^{-1}$.

We will apply the full set of the Planck 2015 likelihood in both
temperature and polarization (referred as PlanckTT, TE,
EE+lowTEB).
%In particular, the main aim is to reducing the
%degeneracy between the tensor-to-scalar ratio and the tilt of
%tensor perturbations.
The Planck likelihood \cite{Adam:2015rua} combined with the
BICEP/Keck data \cite{Array:2015xqh} (BK14) will be used. In all
runs, we also include a prior on the Hubble parameter from the HST
\cite{Scoville:2006vr} and BAO
\cite{Beutler:2011hx,Ross:2014qpa,Anderson:2013zyy}.

Our analysis of combining CMB and interferometers data will be
computed based on the recent CMB data and the upper limits put by
recent aLIGO O1 on stochastic GWs background $\Omega_\mathrm{GW}$.
%Moreover, we also consider combinations of the CMB data and
%the upper limits put by the expected ET and LISA's
%sensitivities.
 In order to perform such analysis, we add a module
computing $\Omega_\mathrm{GW}$ into the {\tt CosmoMC}
\cite{Lewis:2002ah} and the {\tt CAMB} \cite{Lewis:1999bs}.

In Table.\ref{tab:location_of_detectors}, we list our best-fit
results for the parameters set of the lensed-$\Lambda$CDM model
$\{\Omega_bh^2, \Omega_ch^2, 100\theta_\mathrm{MC}, \tau,
\ln(10^{10}A_s), n_s\}$. We see that considering the aLIGO O1 data
 does not significantly alter the results provided by CMB data. Intuitively,
this result is because the GWs only contribute a small fraction of
CMB fluctuations $r\ll 1$.

%\begin{figure}[htbp]
%\includegraphics[scale=2,width=0.75\textwidth]{All.eps}
%\caption{Triangle plot showing the one-dimensional posterior
%distribution of the parameters $(\Omega_bh^2, \Omega_ch^2,
%100\theta_\mathrm{MC}, \tau, \ln[10^{10}A_s], n_s)$ of the
%lensed-$\Lambda$CDM model and their two-dimensional probability
%contours at $68\%$ and $95\%$ CL. "All'' refers to
%\textit{Planck}2015, HST, BAO and BK14 data. }
%\label{fig1constraint}
%\end{figure}

\begin{table}[!bth]
%\begin{ruledtabular}
\begin{tabular}{lcccc}
\hline\hline
 ~ & $\bf All$ & $\bf All+aLIGO$ & $\bf All+LISA$ & $ \bf All+ET$\\
\hline
\ $\Omega_{b}h^2$  & $0.022$ & $0.022$  & $0.022$ & $0.022$\\
\ $\Omega_{c}h^2$  & $0.119$ & $0.119$ &  $0.119$  & $0.118$\\
\ $100\theta_\mathrm{MC}$         & $1.04$ & $1.04$ &  $1.04$ & $1.04$\\
\ $\tau$           & $0.08$ & $0.08$ &  $0.08$ & $0.09$\\
\ $ln(10^{10}A_s)$  & $3.12$ & $3.14$ &  $3.15$ & $3.15$ \\
\ $n_s$            & $0.968$ & $0.971$ &  $0.968$ & $0.968$ \\
\hline\hline
\end{tabular}
%\end{ruledtabular}
\caption{Summary of constraints on the cosmological
6-parameters set $\{\Omega_bh^2, \Omega_ch^2,
100\theta_\mathrm{MC}, \tau, \ln(10^{10}A_s), n_s\}$. ``All''
refers to \textit{Planck}2015, HST, BAO and BK14 data. }
\label{tab:location_of_detectors}
\end{table}

\begin{figure}[htbp]
\includegraphics[scale=2,width=0.5\textwidth]{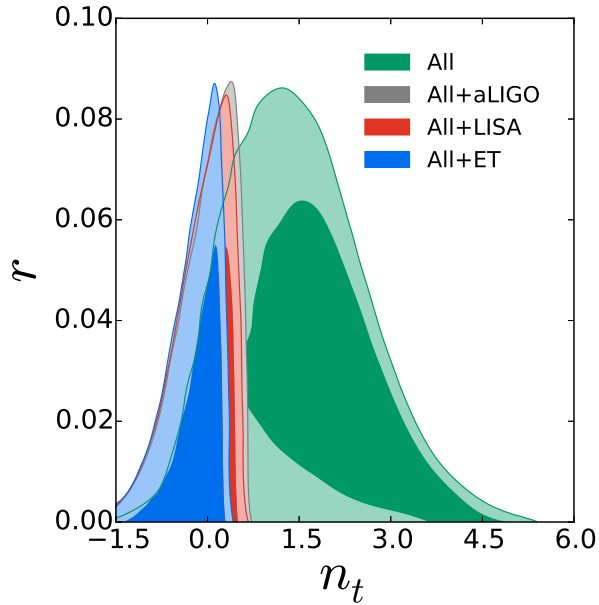}
\caption{Marginalized posterior distributions at 68\% and 95\%
C.L. for the parameters $r$ and $n_T$.
``All'' refers to
\textit{Planck}2015, HST, BAO and BK14 data. }
\label{figconstraint}
\end{figure}

However, the constraint on $r$ and $n_T$ will be affected
importantly, which is plotted in Fig.\ref{figconstraint}. Only CMB
data is used, the posteriors for $n_T$ favor a blue tilt. When we
consider the CMB+aLIGO O1 dataset, the upper limit for $n_T$ 
is cut by a big margin, since a large $n_T$ could bring a detectable 
GWs signal at small scales. The constraint result is presented in 
Table.\ref{table:n_t values}, in which $n_T = 0.016^{+0.614}_{-0.989}$.
 Here, since $k_*$ in (\ref{eq:tensor_pk}) is set
around CMB scale, only the CMB data is sensitive to the GWs
background with $n_T<0$.

 %{\color{red}{and also CMB+ET set and
%CMB+LISA set for forecasts, the upper limit for $n_T$ is cut by a
%big margin, since a large $n_T$ could bring a detectable GWs
%signal at small scales. Our result is presented in
%Table.\ref{table:n_t values}, in which $n_T =
%0.016^{+0.614}_{-0.989}$, $-0.050^{+0.612}_{-0.992}$, $
%-0.184^{+0.534}_{-0.877}$ for above sets, respectively. This
%indicates that the ground- and space-based interferometers are
%able to significantly improve the limits on the tilt $n_T$ of
%primordial GWs. Here, since $k_*$ in (\ref{eq:tensor_pk}) is set
%around CMB scale, only the CMB data is sensitive to the GWs
%background with $n_T<0$.}}

In Ref.\cite{Lasky:2015lej}, Lasky et.al utilized LIGO S5 data
collected in 2009-2010, see also \cite{Huang:2015gka}. In
Ref.\cite{Cabass:2015jwe}, Cabass et.al also performed a combined
analysis of Planck2015+BK14 and aLIGO data. However, the aLIGO
sensitivity they used is $\Omega_\mathrm{GW}\sim 10^{-9}$ at
$f=100$ Hz, which only serves as a forecast, while the current
result of aLIGO O1 is $\Omega_\mathrm{GW}\sim 10^{-8}$, see
Sec.\ref{Sec:SecIIA}.
% at $f\simeq 30$Hz
Thus we provide an updated limit on $n_T$ by using
up-to-date CMB+aLIGO datasets.

\subsection{Forecasts}

In a slightly lower frequency-band, i.e. 1mHz, planning
space-based detector LISA will possibly set
$\Omega_\mathrm{GW}\sim 10^{-14}$
\cite{Bartolo:2016ami, Klein:2015hvg}. The sensitivities of LISA's
different configurations are different. Here, we will use
$\Omega_\mathrm{GW}<\rm{3.72 \times 10^{-13}}$ at $f\simeq
3.8\times10^{-3}$\,{\rm Hz} for our analysis. This upper limit get
from the configuration N2A2M2L6, where N2 presents the best
low-frequency noise level, A2 represents 2 millon kilometers arm
length, 2-years duration of observation is represented by M2, and
L6 corresponds to 6 links.

The ET is a proposed ground-based GWs detector, which belongs to
the third generation detector. It is designed to consist of six
Michelson interferometers, which form an equilateral triangle. The
arm length will be reached $10$ kilometers long. The ET will
possibly put the upper limit for stochastic GWs background to
$\Omega_\mathrm{GW}<\rm{ 10^{-13}}$ at $f\simeq 10$\,{\rm Hz}
\cite{Hild:2010id, Smith:2016jqs}.

We consider the combinations of the CMB data and the upper limits 
put by the expected ET and LISA's sensitivities. From the Table.\ref{tab:location_of_detectors},
we see that the expected ET and LISA's sensitivities does not
significantly alter the cosmological 6-parameters results provided by CMB data. 
yet the constraint on $r$ and $n_T$ is significantly, The forecast result of CMB+ET and 
CMB+LISA are also presented in Table.\ref{table:n_t values}, in which $n_T =
-0.050^{+0.612}_{-0.992}$, $-0.184^{+0.534}_{-0.877}$ for above datasets, respectively.
This indicates that the ground- and space-based interferometers are able to significantly 
improve the limits on the tilt $n_T$ of primordial GWs. In fact, besides these current ground-
and space-based CMB observation, some other measurements can also give
 the constraint on $n_t$. In Ref.\cite{Liu:2015psa}, Liu and Zhao et.al utilized PTA data
to show $n_t$ upper limit, if $r = 0.01$, the optimal gives limit $n_t<0.18$ .

As far as the forecasts are concerned, in 2020s the sensitivity of
aLIGO/Virgo O5 will arrive at $\Omega_\mathrm{GW}\sim 10^{-9}$
\cite{TheLIGOScientific:2016wyq}. However, it cannot improve the
current limit on $n_T$ put by O1 well, and is still weaker than
the limit provided by the future LISA, which will operate in
2030s, since at LISA frequency-band the sensitivity of O5
corresponds to
$\Omega_\mathrm{GW}=(k_{LISA}/k_{aLIGO})^{n_T}\Omega_\mathrm{GW}^{aLIGO}\sim
10^{-10}$, which is still smaller than the LISA's
$\Omega_\mathrm{GW}\sim 10^{-13}$. The ET actually may improve the
limits on $n_T$ slightly better than LISA, since at LISA
frequency-band the sensitivity of ET corresponds to
$\Omega_\mathrm{GW}=(k_{LISA}/k_{ET})^{n_T}\Omega_\mathrm{GW}^{ET}\sim
10^{-14}$. However, here we only used the sensitivity of LISA's
configuration N2A2M2L6, one among LISA's six-link representative
configurations \cite{Bartolo:2016ami}, for our joint analysis. In
all representative configurations of LISA, N2A5M5L6 (5 millon
kilometers arm length and 5-years mission duration) may have
higher sensitivity.

\begin{table*}[htbp!]
\begin{center}
\begin{tabular}{|c|c|c|}
\hline Dataset &\multicolumn{2}{c|}{Parameter} \\ \hline
 & \multicolumn{1}{c|}{\,\,$n_T$\, 95\% limits\,\,} &\multicolumn{1}{c|}{\,\,$r$\, 95\% limits\,\,}\\ \hline
\hline
All &$1.83^{+2.12}_{-2.07}$ &$<0.07$\\
All+aLIGO &$0.016^{+0.614}_{-0.989}$  & $< 0.066$\\
All+LISA & $-0.050^{+0.612}_{-0.992}$&$<0.062$ \\
All+ET    & $-0.184^{+0.534}_{-0.877}$&$<0.061$   \\
%Planck+HST+BAO+BK14+Taiji&0.14 &$-0.029^{+0.549}_{-0.989}$ & 0.025 &$<0.064$ \\
\hline \hline

\end{tabular}
 \caption{Summary of parameter constraints from different
datasets. ``All'' refers to \textit{Planck}2015, HST, BAO and
BK14 data.
 }
 \label{table:n_t values}
\end{center}

\end{table*}

\section{The models confronted with data}
\label{Sec:SecIII}

Generally, the quadratic action of GWs mode $\gamma_{ij}$ is \ba
S^{(2)}_{\gamma}=\int d^4x{a^3 Q_TM_p^2\over8}\lf[
\dot{\gamma}_{ij}^2 -c_T^2{(\partial_k\gamma_{ij})^2\over
a^2}\rt]\,, \label{tensor-action} \ea where $Q_T= f+{2m_4^2\over
M_p^2}>0$, and the propagating speed $c_T^2={f\over  Q_T  }>0$,
see Appendix.\ref{app} for the details of $f$ and $m_4$.

The combination of GWs detectors and CMB data could significantly
intensify the limit on $n_T$, as has been illustrated. Thus the
combined dataset will put tighter constraint on the early universe
models, which produce blue-tilted GWs spectrum ($n_T>0$). Below,
we will confront some of corresponding models, which may be mapped
into EFT, with the interferometers and CMB datasets.

\subsection{The inflation with diminishing $c_T$}

The inflation is the standard paradigm of the early universe.
Based on EFT of inflation, we found that the diminishment of the
propagating speed $c_T$ of GWs during inflation will lead to a
blue-tilted GWs spectrum
\cite{Cai:2016ldn, Cai:2015yza} \be n_T\simeq {p\over
1+p}>0,\label{nT1}\ee where ${\dot c}_T<0$ and $p=-{\dot
c}_T/(H_{inf}c_T)>0$ is assumed to be constant for simplicity, and
the parameter $\epsilon_{inf}\ll 1$ has been neglected. Recently,
it has been proved in \cite{Cai:2016ldn} that the slow-roll
inflation with the diminishing GWs propagating speed (${\dot
c}_T<0$) is disformally dual to the superinflation
\cite{Piao:2004tq, Piao:2007ne}.

The inflation model with massive graviton $m_{graviton}\simeq
H_{inf}$ can also produce a blue-tilted GWs spectrum
$n_T\simeq {\cal O}(1){m^2_{graviton}\over H^2_{inf}}>0$
\cite{Cannone:2014uqa, Bartolo:2015qvr}. In
Ref.\cite{Bartolo:2016ami}, Bartolo et.al. discussed the models
with constant $m_{graviton}$ and $c_T$.
%Higher-order
%curvature corrections to GR also could result in the nontrivial
%GWs spectrum
%\cite{Cai:2015dta}\cite{Cai:2015ipa}\cite{Baumann:2015xxa}.

We will apply the combination of interferometers and CMB data to
put the constraint on the inflation model with ${\dot
c}_T<0$. The Lagrangian in \cite{Cai:2016ldn} is (\ref{action})
with \be f=1,\ee \be c(t)={\dot\phi}^2/2,\qquad \Lambda(t)=V \ee
\be M_2,m_3,{\tilde m}_4=0,\label{mtilde}\ee \be m_4^2(t)=
\left({1\over c_T^2(t)}-1\right){M_p^2/ 2}. \label{m42}\ee The
spectrum of primordial GWs is \cite{Cai:2016ldn} \be P_T\simeq  {2
H_{inf}^2\over M_P^2 c_T\pi^2 } \left({k\over k_*}\right)^{n_T},
\label{nT}\ee with $n_T$ given by (\ref{nT1}), which is
blue-tilted, where $k_* ={(1+p)}{aH_{inf}\over c_T }$. We have
$n_T\simeq p$ for $p\ll 1$ and $n_T\simeq 1$ for $p\gg 1$. Since
$m_4\neq {\tilde m}_4$, our Lagrangian belongs to a subset of
beyond Horndeski theory \cite{Gleyzes:2014dya}.

Here, both the scalar perturbation and the background are
unaffected by ${m}_4(t)$, as has been confirmed in
\cite{Cai:2016ldn}. The background is the slow-roll inflation with
$0<\epsilon\ll 1$, which indicates that the scalar spectrum is
flat with a slightly red tilt and is consistent with the
observations.

\begin{figure}[htbp]
\includegraphics[scale=2,width=0.55\textwidth]{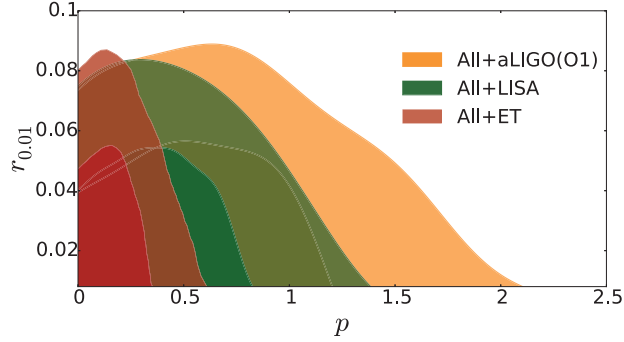}
\caption{Marginalized posterior distributions at 68\% and 95\%
C.L. for the parameters $r$ and $p$. ``All'' refers to
\textit{Planck}2015, HST, BAO and BK14 data. }
\label{figconstraint1}
\end{figure}

\begin{figure}[htbp]
{\includegraphics[width=.46\textwidth]{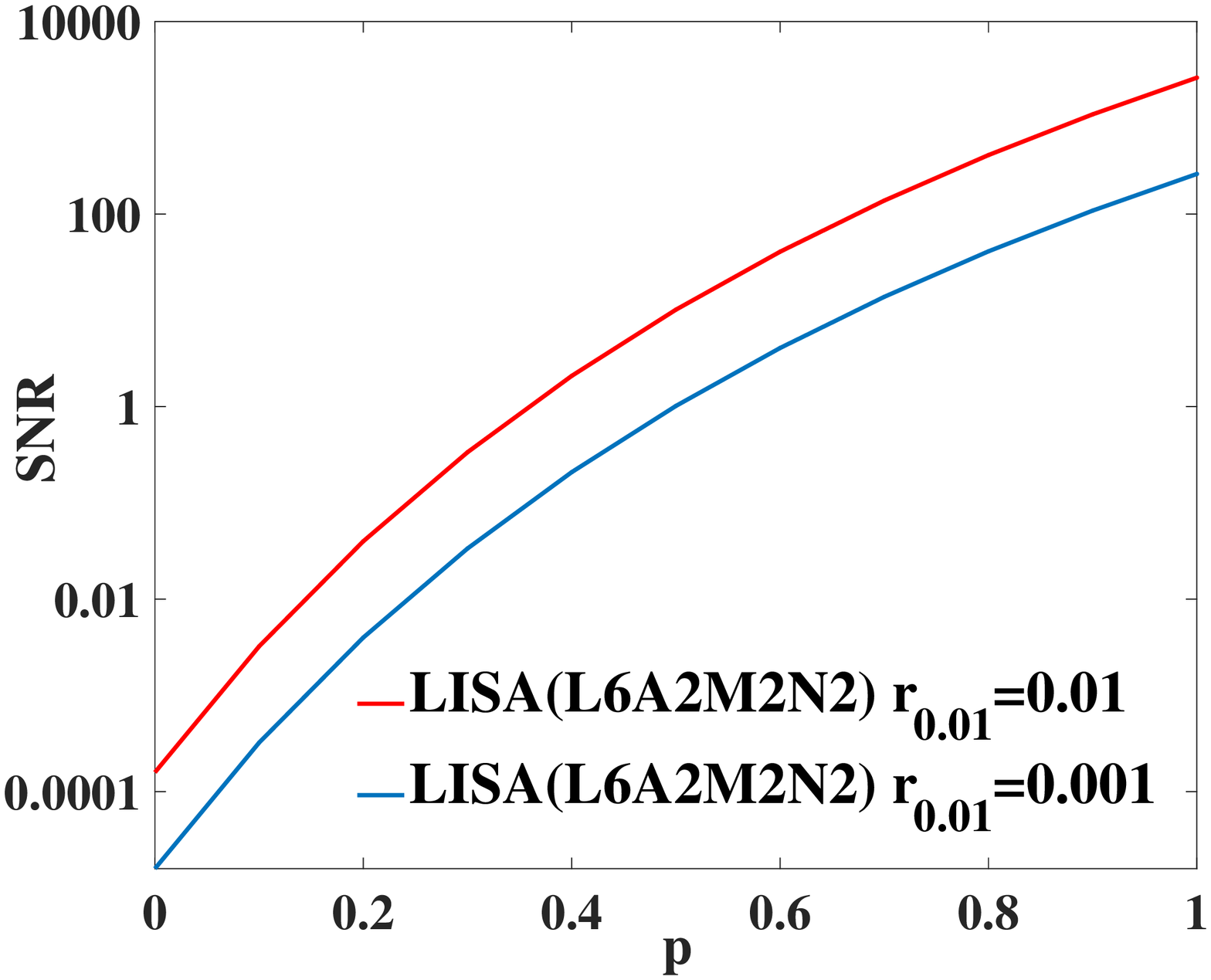} }
%\subfigure[~~Taiji]{\includegraphics[width=.47\textwidth]{r_fc_2yearsTaiji.eps} }
{\includegraphics[width=.46\textwidth]{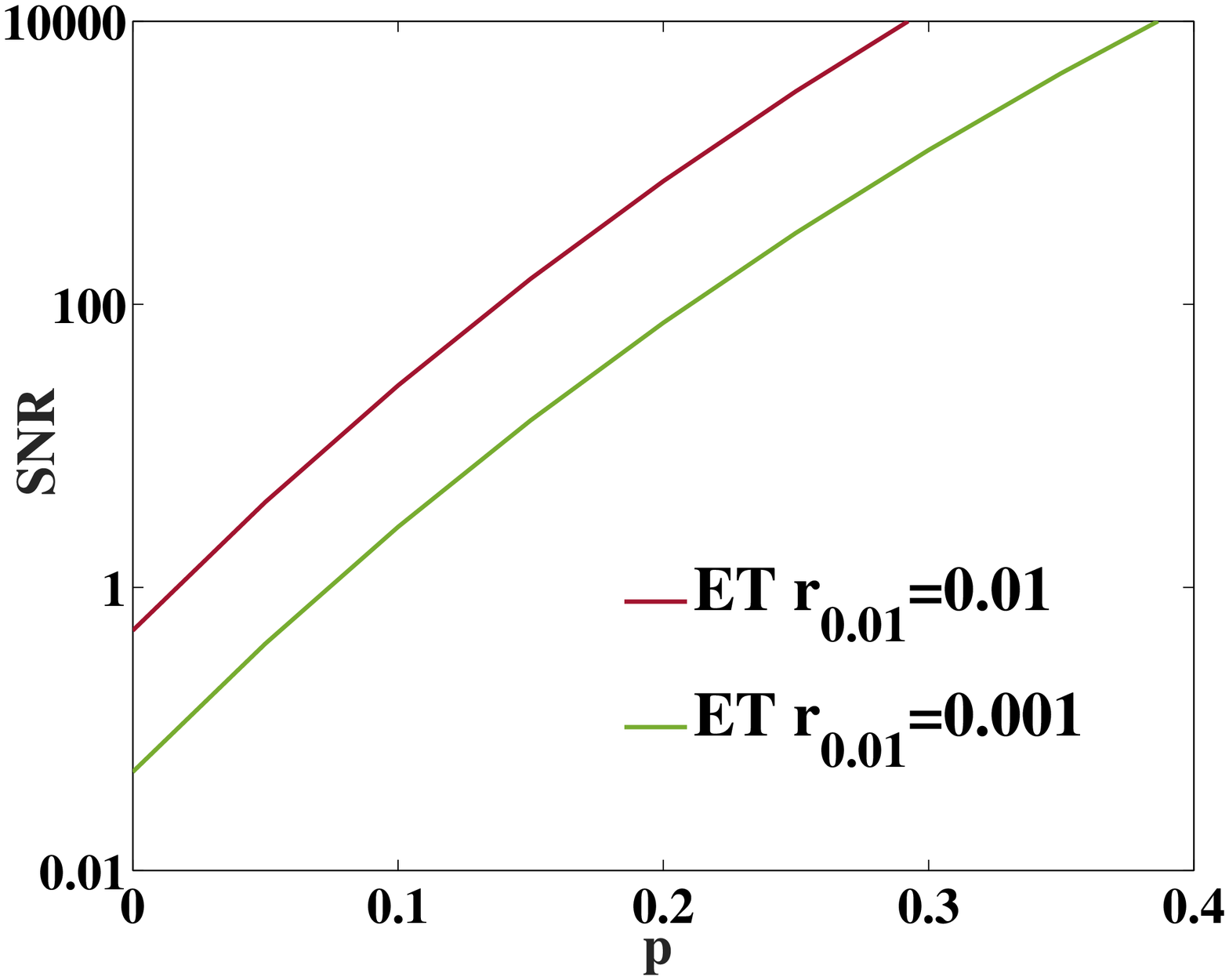} }
%\subfigure[~~Taiji]{\includegraphics[width=.47\textwidth]{nt_fc_2yearsTaiji.eps} }
\caption{The SNRs of LISA configuration N2A2M2L6 and ET with
respect to $p$ for different $r_{0.01}$. The duration of
observation of ET is taken as 1 year. } \label{SNR1}
\end{figure}

We require that after the inflation, $c_T=1$ and GR is recovered.
In Fig.\ref{figconstraint1}, we see that the combination of
Planck2015+BK14 and aLIGO O1 data gives $p=-{\dot
c}_T/(H_{inf}c_T)\lesssim 1.2$ at 68\% C.L. for $r_{0.01}\lesssim
0.01$. In light of (\ref{m42}), this suggests in unit of Hubble
time \be {{\dot m}_4\over H_{inf}m_4}\lesssim 1.2 \ee at 68\% C.L.
for $0<c_T^2\ll 1$. In the future, LISA and ET will provide
stronger limits on $m_4$, which are ${{\dot
m}_4/(H_{inf}m_4)}\lesssim 0.7$ and $0.2$, respectively. We also
calculate the signal-to-noise ratio (SNR) of LISA configuration
N2A2M2L6 and ET used in Fig.\ref{figconstraint1} with respect to
$p$ for different $r_{0.01}$, which is plotted in Fig.\ref{SNR1}.

%It should be mentioned that although $f(t)$ and $m_4(t)$ also
%affect the scalar perturbation, it is always possible to let the
%spectrum of scalar perturbation unchanged by suitably setting the
%behaviors of $M_2,m_3,{\tilde m}_4$.

\subsection{The slow expansion with ${\cal L}_4$}

In Galilean Genesis
\cite{Creminelli:2010ba, Creminelli:2012my} (also slow
expansion scenario \cite{Piao:2003ty, Liu:2011ns}), the
spacetime is flat Minkowski in infinite past and the universe is
slowly expanding, \be a\sim e^{{1\over (-t)^n}}\simeq 1+{1\over
(-t)^n}\ee where $n>0$ and $H\sim {1\over (-t)^{n+1}}$ rapidly
rises, after the slowly expanding phase ends, the universe reheats
and the evolution of hot `big-bang' starts. With the cubic
Galielon, it has been found in \cite{Liu:2011ns} that during the
slow expansion, the scale-invariant adiabatic perturbation may be
produced for $n=4$, see also \cite{Nishi:2015pta}.

The original Genesis scenarios predict a blue-tilted GWs spectrum,
\be P_T\simeq {H^2_{end}\over M_p^2}\left({k\over
k_{end}}\right)^2,\ee The case is similar in ekpyrotic
scenario \cite{Boyle:2003km}(different case see\cite{Ben-Dayan:2016iks}). Here, the pivot scale $k_*$ is
actually $k_*=k_{end}$, where the subscript `end' corresponds to
the quantities at the ending time of the Genesis phase. Thus at
CMB scale $k_{CMB}\sim 0.01\,\mathrm{Mpc}^{-1}$,
 \be r_{0.01}\sim
\left({k_{CMB}\over k_{end}}\right)^2\ll 0.001, \ee is negligibly
small, see \cite{Baumann:2007zm} for a theoretical lower
limit. And at aLIGO scale, $\Omega_{\text{GW}}\sim P_T\sim
({k_{aLIGO}\over k_{end}})^2$ is also far small, unless
$k_{aLIGO}\simeq k_{end}$. Thus the current limit on $n_T$ can
hardly put any constraints for the corresponding models.

%Actually, if we take $k_*\gg k_{CMB}$ for the joint analysis in
%Sec.III, the limit on $n_T$ will be remarkably weaken, see also
%\cite{Bartolo:2016ami}.

%In the slowly evolving phase before $t_{end}$, GR might be
%modified, if so, the scale-invariant GWs spectrum will possibly be
%produced \cite{Piao:2011mq}.

In Ref.\cite{Cai:2016gjd}, based on the Horndeski theory with
${\cal L}_4$, it has been found that in Genesis scenario, the
spectrum of primordial GWs may be flat, and even interestingly, it
also may be blue-tilted with $k_*\sim k_{0.01}$ in
Eq.(\ref{eq:tensor_pk}). Recently, in \cite{Nishi:2016ljg}
Nishi and Kobayashi also have found the similar case in full
Horndeski theory \cite{Kobayashi:2011nu}. Thus the combination of
interferometers and CMB data may put the corresponding bound for
it.

We first briefly review the model proposed in \cite{Cai:2016gjd}.
The Lagrangian is \ba &\,&S=\int d^4x \sqrt{-g}\Big[
{1\over2}e^{4\phi/{\cal M}}X-{1\over 8{\cal M}^8}X^3 -\alpha{\cal
M}^4\,e^{6\phi/{\cal M}}
\\\nn&\,&\qquad\qquad\qquad
+{M_P^2 \over 2}\left(4{\cal M}^8 /X^2+1\right)R- {8 M_P^2 {\cal
M}^8 \over X^3}\left[-\left(\Box\phi\right)^2
+\nabla_{\mu}\nabla_{\nu}\phi\nabla^{\mu}\nabla^{\nu}\phi\right]
\Big]\,, \ea where $X={\nabla_\mu\phi\nabla^{\mu}\phi}$. Mapping
it into EFT, we have \ba &\,& f(t)=1+{4{\cal M}^8
\over X^2}\,,\nn\\
&\,& c(t)={1\over2}e^{4\phi/{\cal M}}X-{3\over 8{\cal M}^8}X^3
+\frac{8 \mathcal{M}^8 M_p^2}{X^3} \left(
   15 H^2 X
   +Y
   \right)\,,
   %%%%%%%%%%%%%%%%%%%%%%%%%%%%%%%%%%%%%%%%%%%%%%%%%%%%%%%%%%%%%
\nn\\&\,& \Lambda(t)=-{ X^3\over 4{\cal M}^8} +\alpha{\cal
M}^4\,e^{6\phi/{\cal M}} +\frac{8 \mathcal{M}^8 M_p^2 }{X^3}
   \left(
   9 H^2 X
   +6 H \dot{\phi} \ddot{\phi}
   -Y\right)\,,
   %%%%%%%%%%%%%%%%%%%%%%%%%%%%%%%%%%%%%%%%%%%%%%%%%%%%%%%%
\nn\\&\,& M_2^4(t)=-{3 X^3\over 4{\cal M}^8} -\frac{4
\mathcal{M}^8 M_p^2 }{X^3} \left(
   90 H^2 X
   +Y
\right)\,,
%%%%%%%%%%%%%%%%%%%%%%%%%%%%%%%%%%%%%%%%%%%%%%%%%%%%%%%%%%%%
\nn\\&\,& m_3^3(t)=\frac{16 \mathcal{M}^8 M_p^2}{X^3}
  \left(
  10 H X+\dot{\phi } \ddot{\phi}
  \right)\,,
  %%%%%%%%%%%%%%%%%%%%%%%%%%%%%%%%%%%%%%%%%%%%%%%%%%%%%%%
\nn\\&\,& m_4^2(t)=\tilde{m}_4^2(t)=\frac{8 \mathcal{M}^8
M_p^2}{X^2}\,, \ea where $Y= 11 H\dot{\phi } \ddot{\phi} +2
\dot{H} X -\dot{\phi } \dddot{\phi} +5 \ddot{\phi}^2 $.

We have the solution \be e^{\phi/{\cal M}}=\left({15\over
4}\right)^{1/4} {1\over {\cal M}(t_*-t)}, \label{ephi}\ee \be
{\dot \phi}={{\cal M}\over (t_*-t)}, \label{dotphi}\ee and
$\alpha={2\over 3\sqrt{15}}$. Thus only one adjustable parameter
${\cal M}$ is left. The background of slow expansion is described
by \be a= a_0 e^{\int Hdt} \simeq a_0(1+{1\over {\cal M}^6
M_P^2}{1\over (t_*-t)^{8}})\simeq a_0, \label{a} \ee where $H\sim
{1\over {\cal M}^6 M_P^2}(t_*-t)^{-9}$, and the condition of slow
expansion is ${\cal M}^6 M_P^2 (t_*-t)^{8}\gg 1$, which suggests
$\epsilon\simeq -{\cal M}^6 M_P^2 (t_*-t)^{8}\ll -1$. Thus
initially the universe is Minkowski. When ${\cal M}^6 M_P^2
(t_*-t)^{8}\simeq 1$, the slowly expanding phase ends. Hereafter,
${X/{\cal M}^4}\gg 1$, GR is recovered, and the hot big-bang
universe will start.

Since $f\sim {m}_4^2/M_p^2 \sim {\cal M}^8/X^2$, we have $Q_T\sim
(t_*-t)^4$ and $c_T^2$ is constant. The GWs horizon $\sim 1/H_T$
is $1/H_T\sim {t_*-t}$. Thus the spectrum of primordial GWs is
\cite{Cai:2016gjd} ${P}_T \simeq ({{\cal M}\over M_P})^{1/2}$.
However, generally we may set $Q_T\sim (t_*-t)^p$, and have \be
{P}_T \simeq \left({{\cal M}\over M_P}\right)^{1/2}\left({k\over
k_*}\right)^{n_T}, \label{PT} \ee where $n_T=4-p$, which is
blue-tilted for $p<4$.

Here, the adiabatic scalar perturbation is strong blue-tilted, $
{P}_s \simeq {{\cal M}\over M_P}({k\over k_*})^{12/5}$ and the
primordial density perturbation responsible for the observations
may be induced by the perturbation of a light scalar field, see \cite{Rubakov:2014jja} for a review,
 which is insensitive to $Q_T$.

\begin{figure}[htbp]
\includegraphics[scale=2,width=0.55\textwidth]{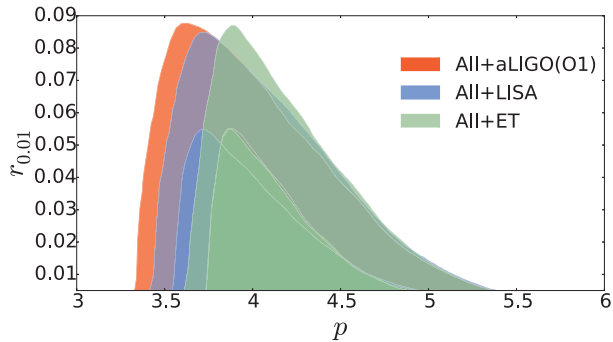}
\caption{Marginalized posterior distributions at 68\% and 95\%
C.L. for the parameters $r$ and $p$. ``All'' refers to
\textit{Planck}2015, HST, BAO and BK14 data. }
\label{figconstraint2}
\end{figure}

In Fig.\ref{figconstraint2}, we see that the combination of
Planck2015+BK14 and aLIGO O1 data gives \be 3.4\lesssim p={d \ln{
Q_T}\over d \ln{H_{T}}}\lesssim 4.6 \ee at 68\% C.L. for
$r_{0.01}\lesssim 0.01$. Thus we could have $r_{0.01}\simeq
0.01-0.001$ at CMB scale, which is detectable for the ongoing CMB
B-mode polarization experiments. Here, since $Q_T\sim f\sim
m_4$, we have $p={d \ln{ f}\over d \ln{H_T}}={d \ln{m_4}\over
d\ln{H_T}}$. In the future, LISA and ET will put stronger bounds,
which are $3.6\lesssim p\lesssim 4.6$ and $3.7\lesssim p\lesssim
4.6$, respectively.

\section{Conclusion}
\label{Sec:SecIV}

The spectrum of primordial GWs, especially $n_T$, carries
significant information about the primordial universe. Though the
primordial GWs remains undetected, combining data at different
frequency-bands is able to put tighter constraint on $n_T$ than
the CMB alone, which helps to deepen our understanding for the
origin of the universe.

We perform a combination of Planck2015+BK14 data and the upper
limit put by recent aLIGO on stochastic GWs background, as well as
the expected ET and space-based LISA sensetivities. Assuming the
spectrum of primordial GWs is described by (\ref{eq:tensor_pk})
with $k_*=0.01 \,\mathrm{Mpc}^{-1}$, we find that with recent CMB+aLIGO O1
dataset, the current limit is $r_{0.01}<0.066$ and
$n_T=0.016^{+0.614}_{-0.989}$ at 95\% C.L., and also for
forecasts, we find that $r_{0.01}<0.062$ and
$n_T=-0.050^{+0.612}_{-0.992}$ for the CMB+LISA set, and
$r_{0.01}<0.061$, $n_T=-0.184^{+0.534}_{-0.877}$ for the CMB+ET
set. Thus at present the CMB+aLIGO O1 dataset has
significantly improved the limit on $n_T$, and it is expected that
in the future LISA and ET would make limits tighter.

We also explore how to confront the models of early universe
scenarios, which produce blue-tilted GWs spectrum, with the
corresponding data. We apply the combined analysis of
interferometers and CMB data to some early universe models, which
may be mapped into EFT. In particular, for the inflation model
with diminishing $c_T$, we find that the current limit is ${{\dot
m}_4/ (H_{inf}m_4)}\lesssim 1.2 $ at 68\% C.L. for
$r_{0.01}\lesssim 0.01$ and $0<c_T^2\ll 1$. In the future, LISA
and ET could put stronger limits on $m_4$, which are ${{\dot
m}_4/(H_{inf}m_4)}\lesssim 0.8$ and $0.3$, respectively.

%furthermore, the detail of constrain the parameters in modified
%gravity, such as gravitational chirality
%theory\cite{Wang:2014abh,Gerbino:2016mqb} by using space-based
%detectors is our next work plan.

\textbf{Acknowledgments}

We thank Sai Wang for his help in program and Bin Hu for
his valuable comments on the earlier manuscript. We acknowledge
the use of \texttt{CAMB} and \texttt{CosmoMC}. This work is supported by NSFC, No.
11222546, 11575188, 11690021, and also supported by the Strategic
Priority Research Program of CAS, No. XDA04075000, XDB23010100.
Z.G.Liu is supported in part by the fifty-seventh batch of China Postdoctoral Fund.

\appendix

\section{EFT and tensor perturbation}
\label{app}

With the ADM line element, we have
\begin{equation}
g_{\mu\nu}=\left(
  \begin{array}{cc}
  N_kN^k-N^2 &  N_j\\
  N_i &  h_{ij}\\
  \end{array}
\right) \,,\qquad
g^{\mu\nu}=\left(
  \begin{array}{cc}
  -N^{-2} &  {N^j\over N^2}\\
  {N^i\over N^2} &  h^{ij}-{N^iN^j\over N^2}\\
  \end{array}
\right) \,,\qquad
\end{equation}
and $\sqrt{-g}=N\sqrt{h}$, where $N_i=h_{ij}N^j$. We can define
the unit one-form tangent vector $n_{\nu}=n_0
(dt/dx^{\mu})=(-N,0,0,0)$ and $n^{\nu}=g^{\mu\nu}n_{\mu} =({1/
N},-{N^i/ N})$, which satisfies $n_{\mu}n^{\mu}=-1$. The induced
3-dimensional metric on the hypersurface is
$H_{\mu\nu}=g_{\mu\nu}+n_{\mu}n_{\nu}$, thus
\begin{equation}
H_{\mu\nu}=\left(
  \begin{array}{cc}
  N_kN^k &  N_j\\
  N_i &  h_{ij}\\
  \end{array}
\right) \,,\qquad
H^{\mu\nu}=\left(
  \begin{array}{cc}
  0 &  0\\
  0 &  h^{ij}\\
  \end{array}
\right) \,.\qquad
\end{equation}
The Ricci scalar is decomposed as \be \label{Ricci}
R=R^{(3)}-K^2+K_{\mu\nu}K^{\mu\nu}+2\nabla_\mu(Kn^\mu-n^\nu\nabla_\nu
n^\mu)~. \ee where $R^{(3)}$ is the induced 3-dimensional Ricci
scalar associated with $H_{\mu\nu}$, and the extrinsic curvature
$K_{\mu\nu}$ on the hypersurface is
$K_{\mu\nu}\equiv{1\over2}{\cal L}_{n}H_{\mu\nu}$ and ${\cal
L}_{n}$ is the Lie derivative with respective to $n^{\mu}$.

Without higher-order spatial derivatives, the EFT of cosmological
perturbation reads \cite{Gleyzes:2013ooa, Piazza:2013coa} \ba
S&=&\int d^4x\sqrt{-g}\Big[ {M_p^2\over2}
f(t)R-\Lambda(t)-c(t)g^{00}
%%%%%%%%%%%%%%%%%%%%%%%%%%%%%%%%%
\nn\\
&\,&+{M_2^4(t)\over2}(\delta g^{00})^2-{m_3^3(t)\over2}\delta
K\delta g^{00} -m_4^2(t)\lf( \delta K^2-\delta K_{\mu\nu}\delta
K^{\mu\nu} \rt) \nn\\ &\,& + {\tilde{m}_4^2(t)\over
2}R^{(3)}\delta g^{00}\Big] +S_m[g_{\mu\nu},\psi_m]\,,
\label{action}\ea where $ \delta g^{00}=g^{00}+1$, $\delta
K_{\mu\nu}=K_{\mu\nu}-H_{\mu\nu}H$, $\delta
K^{\mu\nu}=K^{\mu\nu}-H^{\mu\nu}H$, and $\delta K=\delta
K^{\mu}_{\mu}=K^{\mu}_{\mu}-3H$. The coefficients $(f, c, \Lambda,
M_2, m_3$, $m_4, {\tilde m}_4)$ specify the corresponding
theories. A particular subset $(m_4= {\tilde m}_4)$ of EFT
(\ref{action}) is the Horndeski theory \cite{Horndeski:1974wa}.
$S_m[g_{\mu\nu},\psi_m]$ is the matter part, which is
minimally coupled to the metric $g_{\mu\nu}$.

In the unitary gauge, we set \be
h_{ij}=a^2e^{2\zeta}(e^{\gamma})_{ij},\qquad
\gamma_{ii}=0=\partial_i\gamma_{ij} \,. \ee

 \end{document}